\documentclass[11pt,a4paper]{article}

\usepackage[margin=1.8cm]{geometry}
\usepackage{amsmath}
\usepackage{microtype}
\usepackage{ragged2e}
\usepackage{enumitem}
\usepackage{cite}
\usepackage{array}
\usepackage{rotating}
\usepackage{pdflscape}
\usepackage[colorlinks=true,allcolors=blue]{hyperref}
\usepackage[font=small,labelfont=bf,labelsep=period]{caption}
\usepackage{dblfloatfix}

\newcommand{\affil}[1]{\par\small #1\par\normalsize}
\newcommand{\email}[1]{\par\small\textbf{E-mail:} #1\par\normalsize}
\newcommand{\keywords}[1]{\par\small\textbf{Keywords:} #1\par\normalsize}
\newcommand{\suppdata}[1]{\section*{Data availability statement}#1}
\newcommand{\ack}[1]{\section*{Acknowledgments}#1}
\newcommand{\orcid}[1]{\href{https://orcid.org/#1}{\includegraphics[width=8pt]{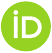}}}

\begin{document}
\justifying

\title{A Frequency-Optimized Optogenetic Study of Network-Level Potentiation in Cortical Cultures on Microelectrode Arrays}
\author{Matteo Dominici$^1$\orcid{0009-0007-5226-7195}, Ilya Auslender$^{1,*}$\orcid{0009-0000-9519-0879}, Clara Zaccaria$^1$\orcid{0000-0001-6660-0779}, Yasaman Heydari$^1$\orcid{0000-0003-2101-0323}\\ and Lorenzo Pavesi$^1$\orcid{0000-0001-7316-6034}}
\date{}
\maketitle

\affil{$^1$Department of Physics, University of Trento, Povo (TN), Italy}

\affil{$^*$Author to whom any correspondence should be addressed.}

\email{ilya.auslender@unitn.it}

\begin{abstract}
\textit{Objective.} Long-term potentiation (LTP) is a fundamental mechanism underlying learning and memory, yet its investigation at the network level \textit{in vitro} remains challenging, particularly when optogenetic stimulation is used. The objective of this work is to develop a robust experimental and analytical framework for inducing and quantifying optogenetically driven LTP in neuronal cultures recorded with microelectrode arrays (MEAs). \textit{Approach.} We first systematically investigate the effect of widefield optogenetic stimulation frequency on evoked neuronal activity, to identify a test-stimulus that reliably probes network responses without inducing activity modulation. By analyzing spike-rate dynamics during repeated stimulation, we characterize frequency-dependent response adaptation consistent with Channelrhodopsin-2 photocycle kinetics. Based on these results, an optimized low-frequency test-stimulus is selected and combined with a spatially confined tetanic optogenetic stimulation to induce LTP. Network responses are quantified using post-stimulus time histograms and a normalized efficacy metric, enabling electrode-wise and network-level analysis of plasticity. \textit{Main results.} Low-frequency optical stimulation ($\leq0.2$\,Hz) preserves stable evoked responses, whereas higher frequencies induce a pronounced sigmoid-like decay in firing rate. Following tetanic stimulation, a subset of electrodes exhibits robust and long-lasting potentiation, persisting for several hours. \textit{Significance.} This work provides a systematic methodology for studying activity-dependent plasticity in optogenetically driven neuronal networks.

\end{abstract}

\section*{Introduction}

Understanding how neuronal networks process information and modify their connectivity in response to activity remains one of the central challenges in neuroscience. Since the seminal work of Hodgkin and Huxley \cite{hodgkin&huxley}, which established a quantitative description of the ionic mechanisms underlying action potential generation, the study of neuronal dynamics has increasingly benefited from interdisciplinary approaches combining experimental neuroscience, physics, and engineering. Despite significant progress, many mechanisms governing collective network behavior and long-term synaptic modifications are still only partially understood.
\\
Simplified neuronal cultures provide a powerful and controllable experimental model for investigating network dynamics \cite{massobrio2015vitro,pozzi2020evaluation}. Compared to \textit{in vivo} systems, neuronal culture networks offer precise control over cellular phenotypes, connectivity, and stimulation parameters, enabling the isolation of specific dynamical processes. These advantages have motivated the development of experimental platforms capable of simultaneously stimulating and recording neuronal activity across multiple spatial and temporal scales. Microelectrode arrays (MEAs, figure \ref{fig:prima_figura}(A)-(B)) are a well-established technology for long-term, non-invasive extracellular recordings, allowing parallel monitoring and electrical stimulation of neuronal networks \cite{thomas1972miniature,gross1977new, pine1980recording,hofmann2006long,obien2015revealing}. However, electrical stimulation is constrained by electrode geometry and lacks cell-type specificity. Optogenetics overcomes these limitations by enabling spatially and temporally precise, cell-type-specific control of neuronal activity through patterned illumination \cite{deisseroth2011optogenetics}. Among the various light patterning techniques, digital light projection (DLP) systems allow the generation of arbitrary light patterns, enabling flexible stimulation and facilitating the investigation of distributed network responses \cite{bhatia2020patterned,jung2017digital,auslender2025decoding, zaccaria2026}.
\\
A key phenomenon in activity-dependent network reorganization is communication strengthening, and in particular long-term potentiation (LTP), which is widely regarded as a cellular substrate of learning and memory \cite{bliss1973long,ltp_1,ltp_2}. \textit{In vitro}, LTP is typically investigated by comparing evoked responses before and after high-frequency or tetanic stimulation protocols. More precisely, a common stimulation protocol follows the scheme: test-stimulus $\rightarrow$ tetanic stimulus $\rightarrow$ test-stimulus \cite{ruaro2005toward, chiappalone2008network,zhang2020familiarity}.
\\
MEA-based recordings are particularly well suited for this purpose, as they enable the assessment of dynamic changes across the entire network over extended time periods. The integration of optogenetic stimulation with MEA recordings, therefore, represents a promising approach for investigating long-term potentiation at the network level, combining precise control of stimulation with large-scale monitoring of network activity \cite{zhang2020familiarity}.

\begin{table*}
\caption{Overview of plasticity-inducing protocols used in literature on cortical cultures on MEAs. We highlight the stimulation approach (electrical or optogenetic), the spatial patterns and frequency $f_0$ used for the test-stimuli, as well as the maximum post-induction time ($T_{max}$) reported in each study. Additional columns concerning spatial and timing settings for tetanus/induction stimuli are added. Reference articles are presented in chronological order.}
\centering
\renewcommand{\arraystretch}{1.5}

\begin{tabular}{@{}
p{1.8cm}
p{3.5cm}
p{1.8cm}
p{3.5cm}
p{3.0cm}
p{0.9cm}
p{0.5cm}
}

\hline
Stimulation & Test-stimulus & $f_0$ & Tetanus/Induction stimulus & Tetanus timing & $T_{max}$ & Ref. \\
\hline
\hline

Electrical &
Single MEA electrode or 5 electrodes simultaneously &
Single pulse or pulses at $1$\,kHz &
Same electrodes of the test-stimulus & Trains of 20 pulses at 20\,Hz, repeated 5–10× at
10–15\,s intervals &
20\,min & \cite{maeda1998modification}\\

Electrical &
Single electrode at a time, sequentially over all 64 electrodes &
Single pulse per electrode, 1 every 3 s &
Single electrode & Trains of 10 pulses at 20\,Hz, repeated 20× at 5\,s intervals &
30\,min & \cite{jimbo1999simultaneous}\\

Electrical &
Fixed electrode pair &
0.3--1\,Hz&
Test-stimuli served as induction stimuli & --- & 60\,ms & \cite{shahaf2001learning}\\

Electrical &
Spatial patterns involving $>$10 electrodes (in a $L$-shape) &
0.5\,Hz &
Same electrodes of the test-stimulus & Trains of 100 pulses at 250\,Hz, repeated 40× at 2\,s intervals &
1\,h & \cite{ruaro2005toward}\\

Electrical & Sequential single-electrode probing & $0.33$\,Hz & Single electrode or small clusters & Trains of 20 pulses at 100\,Hz, repeated 150× at 6\,s intervals & 1\,h & \cite{wagenaar2006searching}\\

Electrical & None; plasticity evaluated on spontaneous network activity & --- & Two electrodes simultaneously & Continuous 20\,Hz stimulation for 15\,min & 3\,h & \cite{madhavan2007plasticity}\\
 
Electrical & 6–8 electrodes, stimulated sequentially & 0.2\,Hz & Subset of test-stimulus electrodes & 20 bursts at 0.2\,Hz, each burst consisting of 11 pulses at 20\,Hz & 2.5\,h & \cite{chiappalone2008network}\\

Optogenetic &
None; plasticity assessed from spontaneous activity &
--- &
Widefield optical tetanus & Trains of pulses at 40\,Hz lasting 1\,s, repeated every 10\,s for 10\,min &
1\,h & \cite{lignani2013long}\\

Optogenetic & None; plasticity assessed from spontaneous activity & --- & Widefield optical stimulation & 40 light pulses lasting 1\,s, delivered at 0.5\,Hz & 12\,min & \cite{el2013optogenetic}\\

Optogenetic &
Pattern-based optical stimuli (complex images) &
0.1\,Hz &
Same spatial pattern of test-stimulus & Trains of pulses at 50\,Hz (10\,ms on / 10\,ms off), lasting 1\,s, repeated 60× at 9\,s intervals &
24\,h & \cite{zhang2020familiarity}\\

\hline
\end{tabular}
\label{other_articles}
\end{table*}

To our knowledge, experimental protocols for inducing LTP on MEAs in neuronal cultures are considerably more established for electrical stimulation than for optogenetic stimulation, for which only a limited number of studies are currently available (see table \ref{other_articles}). Moreover, both in the case of electrical or optical stimulation, none of the existing works appears to address two critical aspects (see table \ref{other_articles}):
\begin{itemize}
    \item a systematic determination of the optimal frequency for the test-stimulus (the most proximate work can be found in \cite{wagenaar2005controlling}, where a frequency study to suppress spontaneous bursting in cortical cultures is presented, to facilitate more reliable and predictable outcomes in tetanization experiments, exclusively for electrical stimulation);
    \item the combination of a widefield test-stimulus with a spatially confined tetanic stimulation.
\end{itemize}
The test-stimulus should be properly designed not to induce any lasting modification of network activity, which must instead be selectively elicited by the tetanic protocol. A systematic study on the test-stimulus frequency is thus needed for reliable experiments. Moreover, we chose the test-stimulus to be of a widefield type, illuminating the highest number of cells possible through the system’s objective. This was motivated by the interest in probing the global functional state of the network reliably. By activating a large fraction of the neuronal culture simultaneously, widefield illumination provides a robust and reproducible measure of the evoked network response, reducing variability associated with local fluctuations in excitability or connectivity \cite{wagenaar2006searching}. Here, the optogenetic approach provides an advantage since it avoids the artifacts associated with electrical stimulation, allowing for the use of a maximized number of recording electrodes. For the induction phase, restricting the tetanic stimulation to a limited set of electrodes allows targeted engagement of specific microcircuits rather than the entire network. The localized induction stimulus makes it possible to subsequently probe, using widefield test-stimulation, how modifications at the circuit level propagate and influence the global network response.
\\
In this work, we present an experimental and analytical framework for studying optogenetically induced LTP \textit{in vitro} using MEA recordings. First of all, different widefield stimulation frequencies are applied to identify a proper test-stimulus that reliably evokes neuronal responses while minimizing activity-dependent modulation. Building on these results, we then apply a protocol to induce and quantify LTP at the network level. This approach enables a systematic characterization of both the temporal evolution and the spatial distribution of plastic changes, providing insights into the mechanisms underlying activity-dependent modulation of neuronal networks.

\section{Materials and Methods}

\subsection{Cultures on MEAs}\label{sec:cells}
\begin{figure*}[h]
\centering
\fbox{
  \begin{minipage}{1.0\linewidth}
    \centering
    \includegraphics[width=\linewidth]{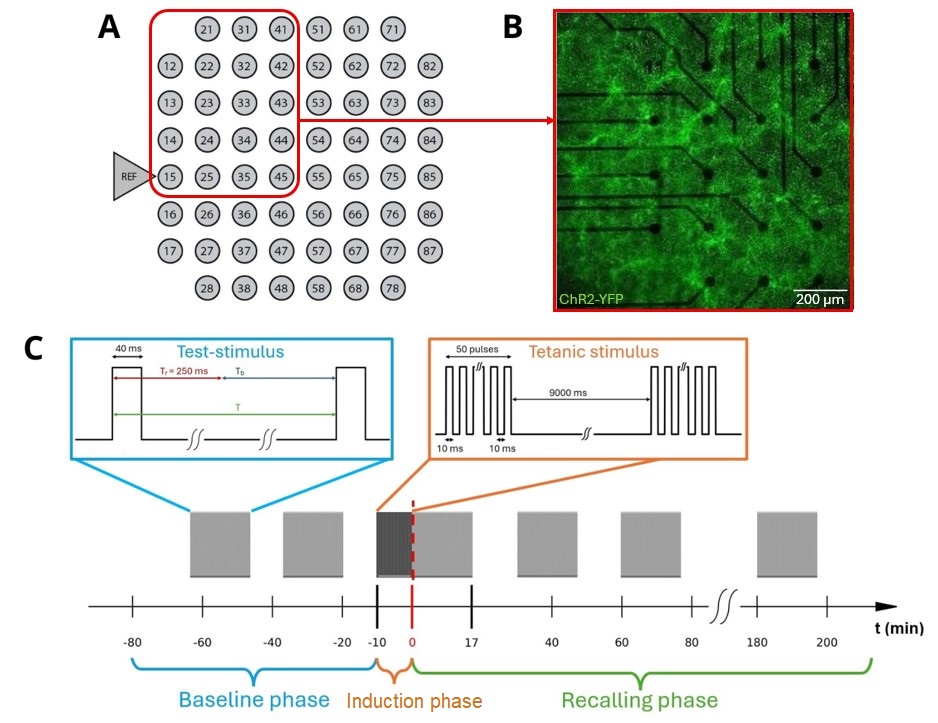}

    \caption{(A): Configuration of the 60 recording electrodes in the MEA chip 60MEA200/30iR-Ti (MCS). The numbering of the MEA electrodes arranged in the 8×8 grid follows the conventional scheme used for square arrays. In this system, the first digit identifies the column number, while the second digit corresponds to the row number. This systematic notation allows for an unambiguous localization of each electrode within the array and ensures consistency between experimental setups and data analysis. Position 15 is occupied by a larger internal reference electrode. (B): Acquisition of YFP image to confirm the expression of ChR2 in the culture. (C): Stimulation protocol for LTP experiments. In this representation the test-stimulus is chosen of frequency $0.2$\,Hz ($\sim$ 17\,min of stimulation, see table \ref{tab1}). During the ‘Recalling phase' the test-stimulus is repeated at time-steps 0, 30, 60, 90, 120, 180 (min) after the conclusion of the tetanic stimulation. The blue box shows the time windows $T_r$ and $T_b$ for the analysis of the spike rate (clarified in section \ref{sec:WF_results}), whilst the orange box depicts the stimulation time-pattern used for the tetanus.}
    \label{fig:prima_figura}
  \end{minipage}
}
\end{figure*}

Cortical neuronal cultures were prepared from wild-type C57BL/6 mouse embryos at embryonic day 17–18 (E17–18). Dissected tissues were enzymatically dissociated by incubation in 0.25\% trypsin–EDTA for 20\,min, followed by enzyme inactivation with high-glucose DMEM supplemented with 10\% fetal bovine serum and penicillin – streptomycin. The resulting cell suspension was filtered, counted, centrifuged (1900\,rpm, 5\,min), and resuspended in seeding medium composed of Neurobasal, 10\% FBS, and penicillin–streptomycin. A total of 136,000 cells in 80\,µL were seeded onto each MEA chip (see section \ref{sec:MEAsetup}) and incubated for 2\,h before addition of feeding medium (Neurobasal supplemented with B27, GlutaMAX, and penicillin–streptomycin). Half of the medium was replaced every $3$-$4$ days. To enable optogenetic stimulation, cultures were transduced at 3 days in vitro (DIV) with an AAV5 vector encoding ChR2(H134R)-EYFP under the human synapsin promoter (pAAV-hSyn-hChR2(H134R)-EYFP), at a titer of 4.6 × 10$^6$ genome copies per chip. Expression of the optogenetic construct was confirmed by YFP fluorescence approximately 7 days post-transduction (figure \ref{fig:prima_figura}(B)). Neuronal activity was subsequently monitored during light stimulation. Experiments were performed between 14 and 20 DIVs.

\subsection{MEA recording system}\label{sec:MEAsetup}

Extracellular electrophysiological recordings were performed using the MEA2100-Mini system supplied by Multi Channel Systems GmbH (MCS), Reutlingen, Germany. The system consists of a compact headstage (HS), a signal collection unit (SCU) and an interface board (IFB), which is connected to a computer. The recording is operated via Multi Channel Suite software (MCS). The headstage provides 60 recording channels with on-board amplification and 24-bit digitization, allowing high signal-to-noise recordings and stable long-term operation inside an incubator. During experiments, cultures were maintained in an incubation system under controlled physiological conditions at 37\,°C and 5\% CO$_2$, with temperature and gas composition regulated by external control units (Okolab). Sterility was maintained with the so called MEA-MEM culture chamber (MCS), integrating a transparent semipermeable Teflon membrane \cite{potter2001new}. The cultures were plated on 60MEA200/30iR-Ti microelectrode array (MEA) chips, composed of 60 planar microelectrodes (TiN/SiN; $30$\,µm electrode diameter; $200$\,µm inter-electrode spacing) arranged in an 8 × 8 grid with the four corner electrodes omitted (figure \ref{fig:prima_figura}(A)). The electrodes are in a glass substrate, which exhibits the serial number of the chip. The chip is equipped with an internal reference electrode (iR), which can be directly connected to the amplifier ground. The center of each MEA chip was double coated with poly-D-lysine and laminin before plating cells on them.
\\
Recorded signals were amplified and digitized at a sampling rate of $20$\,kHz. Data monitoring, acquisition, and storage were performed using the Multi Channel Suite software (MCS). In our work, data analysis was not performed using the proprietary Analyzer module; instead, raw signals were exported and processed offline using custom MATLAB routines. Extracellular signals were processed by a fourth-order Bessel bandpass filter of $300$-$3000$\,Hz to isolate extracellular action potentials from local field potentials \cite{obien2015revealing,belitski2008low,quiroga2009real,zhang2020familiarity}.
The use of this filter also allows for the elimination of an artifact in the raw signal due to photoelectric and photo-thermal effects on the electrodes \cite{wagenaar2002real}.

\subsection{Optical stimulation system}\label{sec:OPTOsetup}
The experiments were performed using an upright confocal microscopy system supplied by Crisel Instruments Company, equipped with an X-Light V2 high-performance spinning-disk confocal module and a Prime Back-Side Illuminated (BSI) scientific CMOS camera (2048 × 2048 pixels). The system was operated in widefield and confocal modalities. A Digital Light Processing (DLP) system (DLP E4500) was optically aligned to the epi-fluorescence rear port of the microscope. The DLP unit comprises three LEDs, an optical module, a WXGA Digital Micromirror Device (DMD; 912 × 1140 mirrors), and a driver board. In this work, the blue LED ($460 \pm 14$ nm) was used, delivering a maximum output power of $600$\,mW. Patterned illumination was directed to the sample through a dichroic mirror (Chroma T505lpxr-UF1), which reflects wavelengths below $505$\,nm while transmitting longer wavelengths for fluorescence detection \cite{zaccaria2026}. Illumination patterns were projected using 8-bit images to ensure uniform light distribution at the sample plane, while LED currents were adjusted via the system control software to modulate the excitation intensity. Temporal control of the projected illumination was achieved via external triggering using a waveform generator (Rigol DG2052), enabling the generation of pulse trains required for the experiments. A 10x objective was used (LMPLFLN10XLWD, NA 0.25, WD 21mm). Light intensity at the sample plane was $2.6\,$mW/mm$^2$. This value was obtained by measuring light power with an Ophir power meter $-$ connected to a PD300-SH silicon photodiode (Ophir) $-$ under the microscope objective, normalizing over the illuminated area, and accounting for the attenuation due to the incubation system's window and the Teflon membrane. 

\subsection{Experimental protocol}
Before starting experiments, a MEA chip culture is placed inside the headstage and translated under the microscope. Cultures were allowed to equilibrate for $\sim10$\,min to ensure that neuronal activity was stable and not affected by the transfer. Transmitted-light images were acquired to verify correct alignment of the objective with the MEA chip, while YFP fluorescence imaging was used to confirm ChR2 expression (figure \ref{fig:prima_figura}(B)). 

\subsubsection{Establishing test-stimulus frequency} \label{sec:WF_exp}

\begin{table*}

    \caption{Experimental procedure for widefield optogenetic stimulation. The table summarizes the sequence of stimulation performed for each culture, focusing attention on the settings used for each of the 5 chosen frequencies. $T$ corresponds to $1/f_0$, whilst $T_{stim}$ indicates the total stimulation time. During data analysis, the period $T$ is divided into two time windows $T_r$ and $T_b$, as will be discussed in section \ref{sec:WF_results}. The last column indicates the resting time introduced between the conclusion of a measure and the start of the measure of the next frequency.}
    \centering
    \begin{tabular}{c c c c c c c}
    \hline
    $f_0$ (Hz) & $T$ (ms) & $T_r$ (ms) & $T_b$ (ms) & Pulses & $T_{stim}$ & Time of rest (min)\\
    \hline
    \hline
    0.1 & 10000 & 250 & 9750 & 200 & 2000\,s ($\sim$ 33.33\,min) & $\sim$ 15 \\
    0.2 & 5000 & 250 & 4750 & 200 & 1000\,s ($\sim$ 16.67\,min) & $\sim$ 25 \\
    0.5 & 2000 & 250 & 1750 & 200 & 400\,s ($\sim$ 6.67\,min) & $\sim$ 40 \\
    1.0 & 1000 & 250 & 750 & 200 & 200\,s ($\sim$ 3.33\,min) & $\sim$ 60 \\
    2.0 & 500 & 250 & 250 & 200 & 100\,s ($\sim$ 1.67\,min) & $-$ \\
    \hline
    \end{tabular}
    \label{tab1}

\end{table*}

Baseline spontaneous network activity was recorded for $10$\,min using the MEA acquisition software. Optical stimulation was delivered using the DLP system in a widefield configuration to achieve global network activation. Light pulse sequences were generated using a custom MATLAB script, which controlled the waveform generator driving the DLP. 
Stimulation consisted of $40$\,ms light pulses (as in \cite{zhang2020familiarity}) delivered at a specific frequency $f_0$ for a total of 200 pulses. For all the cultures under analysis, we decided to investigate five frequencies $f_0$ ($0.1$, $0.2$, $0.5$, $1.0$, and $2.0$\,Hz), starting from the lowest one, while keeping pulse duration and pulse number constant (table \ref{tab1}). Between successive stimulation blocks, resting intervals were included to allow recovery of spontaneous activity and to minimize cumulative effects (table \ref{tab1}). Following the final stimulation sequence, spontaneous neuronal activity was recorded again for $10$ min. During this analysis, a total of $8$ cultures were analyzed.

\subsubsection{LTP experiments}\label{sec:LTP_exp}
Once the optimal test-stimulus frequency ($f_{TS}$) is established, LTP experiments are performed.
The experimental protocol we used for the induction of LTP \textit{in vitro} consists of the following steps (figure \ref{fig:prima_figura}(C)):
\begin{enumerate}[label=\alph*)]
    \item \textit{Baseline phase}
        \begin{enumerate}[label=\arabic*)]
             \item Recording of spontaneous activity ($\sim 20$ min). This step was necessary at the beginning of each experiment to assess the level of activity of the network (i.e., the presence of spikes/bursts);
             \item Test-stimulus 1 (or \textit{Control 1}) $-$ first stimulation of the culture with a test-stimulus of widefield type. The stimulation pattern is identical to the one described in section \ref{sec:WF_exp} (pulse duration $40$ ms, $200$ pulses), but with a pulse period equal to $1/f_{TS}$;
             \item Test-stimulus 2 (or \textit{Control 2}) $-$ ‘baseline’ condition. $10$ minutes after the conclusion of \textit{Control 1}, the test-stimulus is repeated, to assess the stability of the network response;
        \end{enumerate}
    \item \textit{Induction phase}
        \begin{enumerate}[label=\arabic*), start = 4]
             \item Tetanic stimulation (10 min). $10$ minutes after the conclusion of step 3, a high frequency stimulation is applied to two selected electrodes. 
             Their choice is guided by three criteria: (i) strong and reliable evoked responses during the baseline phase, (ii) presence within the field of view of the microscope, and (iii) physical proximity to each other (to increase the probability that they share connections).         
             The stimulation pattern is the same $-$ in terms of frequency and timing $-$ of the one used in \cite{zhang2020familiarity}: it is characterized by $60$ trains of $50$ pulses, with single pulse's on and off times both equal to $10$\,ms (see orange box in figure \ref{fig:prima_figura}(C));
        \end{enumerate}
    \item \textit{Recalling phase}
        \begin{enumerate}[label=\arabic*), start = 5]
             \item ‘Recalling’ step. Step 2 is repeated to assess the level of increase or decrease of the network response; 
             \item Spontaneous activity recording ($10$\,min), starting at the conclusion of step 5;
             \item Steps 5+6 are then repeated at 30, 60, 90, 120, and 180 minutes after the moment of conclusion of tetanic stimulation.            
        \end{enumerate}
\end{enumerate}
The entire stimulation protocol was applied on a total of $7$ cultures.

\subsection{Data processing techniques}
\subsubsection{Spike detection (PTSD)}\label{sec:PTSD}
Extracellular spike detection was performed using the Precision Timing Spike Detection (PTSD) algorithm \cite{maccione2009novel}. This method offers an effective balance between detection accuracy and computational efficiency, making it well suited for large-scale neural data analysis. Spikes were detected using Relative Maximum/Minimum (RMM) identification. For each recording channel, a differential threshold (DT) was defined as a multiple of the standard deviation of the background noise. A spike was identified when the voltage difference between consecutive local extrema within a peak lifetime period (PLP) exceeded the DT. A refractory period (RP) was enforced to prevent multiple detections of the same event. Spike timing was assigned to the sample with the largest absolute amplitude within the detected peak, with boundary cases refined using a short overshoot interval \cite{maccione2009novel}.
\\
In our study, the following parameter values were used: PLP = $1.5$\,ms, RP = $1.0$\,ms, and overshoot = $0.5$\,ms. The differential threshold is set at $7$ times the standard deviation of the noise \cite{chiappalone2005burst,maccione2009novel}.

\subsubsection{Spike rate}\label{sec:rate}
Once spikes have been identified, we have characterized the neuronal activity by means of the firing rate (FR). According to E.\,Adrian’s classical definition \cite{adrian1928physical, bologna2010investigating}, the firing rate corresponds to the number of spikes observed in a relatively large temporal window. More precisely:
\begin{equation}\label{eq:FR_def}
\text{FR} = \dfrac{N}{T}
\end{equation}
where $T$ denotes the total duration of the recording and $N$ is the number of detected spikes in $T$.

\subsubsection{PSTH histograms}\label{sec:PSTH}
To investigate the neuronal activity evoked by stimulation, for each MEA electrode, we computed the Post-Stimulus Time Histograms (PSTH), which is mathematically defined as the sum of all spike trains normalized by the number of stimulus presentations \cite{bologna2010investigating, PSTH_1, PSTH_2}. 
In practice, the time axis is discretized into small intervals of amplitude $\Delta \tau$, and spikes falling within each bin are counted to build the histogram.
Typically, spike counts are normalized not only by the number of stimuli but also by the bin width $\Delta \tau$, yielding firing rates in Hz \cite{rieke1999spikes}:
\begin{equation}
    \text{PSTH}(t,ch) = \frac{1}{\Delta \tau \cdot N_{stim}}\sum^{N_{stim}}_{i=1}N_i(t,ch)
    \label{eq:PSTH}
\end{equation}
where $N_i$ is the total number of spikes falling in the window $\Delta \tau$, whereas $N_{stim}$ is the total number of stimulation events.
\\
In our work, PSTHs were computed over 250\,ms post-stimulus windows using 5\,ms bins. 

\subsubsection{Efficacy}\label{sec:efficacy}
To quantify changes in evoked activity after LTP induction, we compared post-tetanic responses to baseline recordings according to the following method. The areas of the PSTHs measured during \textit{Control 1} and \textit{Control 2} ($Area_{C1}$ and $Area_{C2}$, respectively) were averaged to define a baseline reference value $Area_{BC}$ for each electrode. For each post-tetanus stimulation, the relative change in the evoked response was expressed as a normalized delta value $\Delta A$, also referred to as \textit{efficacy}:
\begin{equation}\label{eq:delta_LTP} 
    \Delta A(ch,t) = \frac{Area_{post}(ch,t) - Area_{BC}(ch)}{Area_{BC}(ch)}
\end{equation} 
where $Area_{post}(ch,t)$ is the PSTH area for an electrode \textit{ch} at the post-tetanus time \textit{t}. This provides a measure of potentiation (or depression) that is independent of the absolute firing rate of the electrode \textit{ch}.

\subsubsection{Potentiation Index}\label{sec:PI}
The potentiation index (PI) is related to changes in the PSTH area (i.e., number of evoked spikes) for each recording channel in two experimental conditions: pre- and post-tetanus-based stimulus \cite{chiappalone2008network}. An electrode has a ‘potentiated response’ at a specific post-tetanus time \textit{t} if it has an increase of the PSTH area to a value equal to or higher than a settled threshold ($S$). More specifically, in our analysis the value $\text{PI}(t)$ is defined as the fraction:
\begin{equation}\label{eq:PI_index} 
    \text{PI}(t) = \frac{N_{thr}(t)}{N_{active}}
\end{equation} 
where $N_{thr}(t)$ is the number of electrodes of the MEA chip that have $\Delta A (t) \ge 1.2\times S$, whereas $N_{active}$ is the number of active electrodes in the chosen MEA chip. The precise choice of the threshold $S$ will be discussed in section \ref{sec:LTP_results}.

\section{Results}

To identify the optimal test-stimulus frequency, we characterized the neuronal response to repeated widefield stimulation at different frequencies, focusing on the temporal evolution of firing rates and on frequency-dependent adaptation effects. The optimal test stimulus is indeed supposed not affect the activity of the network, preserving stable neuronal responses \cite{maeda1998modification,jimbo1999simultaneous,shahaf2001learning,ruaro2005toward,wagenaar2006searching,madhavan2007plasticity,chiappalone2008network,lignani2013long,el2013optogenetic,zhang2020familiarity}.\\
We then established an LTP induction protocol to monitor network activity for up to three hours, characterizing both the activity changes and temporal persistence of potentiation across the MEA. The aim is to evaluate if our setup and stimulation protocol are capable of inducing and quantifying long-term potentiation at the network level. In our experiments, we decided to apply tetanic stimulation only to two nearby, responsive electrodes in order to induce plasticity in a controlled and spatially localized manner. This configuration favors the activation of strongly interconnected neuronal subpopulations, increasing the effectiveness of high-frequency stimulation in triggering potentiation while avoiding widespread network saturation \cite{jimbo1999simultaneous,ruaro2005toward}. Restricting the tetanus to a small region of the MEA enables the propagation of plasticity-related effects from the stimulated sites to the rest of the network \cite{madhavan2007plasticity,chiappalone2008network}.

\subsection{Test-stimulus investigation}\label{sec:WF_results}
\begin{figure*}[h!]
\centering
\fbox{
  \begin{minipage}{1.0\linewidth}
    \centering
    \includegraphics[width=\linewidth]{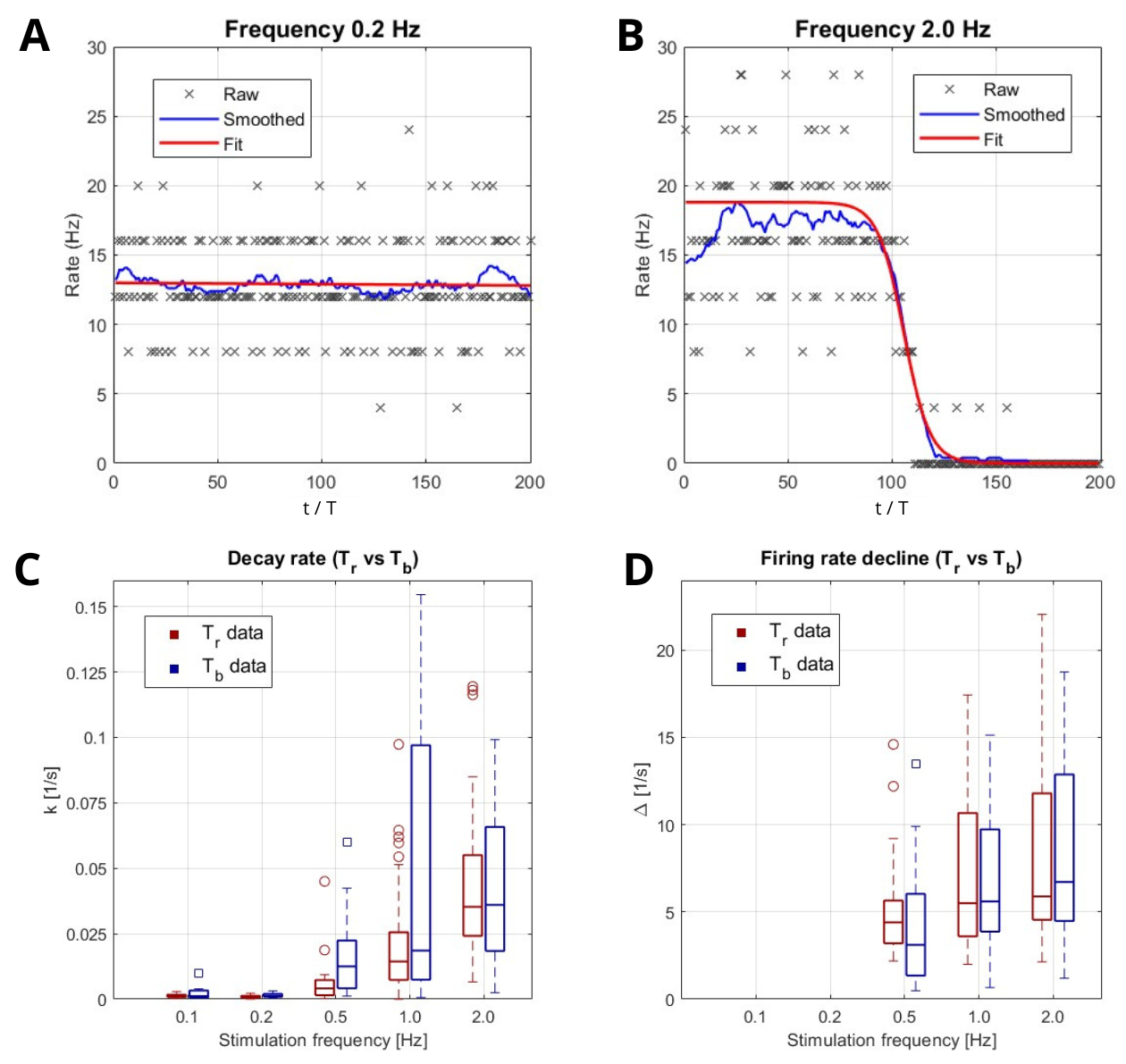}

    \caption{Main results for the test-stimulus. (A)-(B):  Examples of spike rate's evolution for the frequencies $0.2$\,Hz and $2.0$\,Hz for the same electrode (window $T_r$). On the x-axis, $t/T$ is the fraction between the stimulation time and the pulse period $T$ and corresponds to the identification number of each of the total 200 pulses. (C): Evolution of the average decay rate $k$ of the fitting sigmoid, as a function of the stimulation frequency $f_0$ used. (D): Evolution of the average firing rate decline $\Delta$, as a function of the stimulation frequency $f_0$ used. The data corresponding to the frequencies $0.1$ and $0.2$\,Hz are omitted, since sigmoidal fitting becomes difficult for the almost flat behavior.}
    \label{fig:results_teststimulus}
  \end{minipage}
}
\end{figure*}

To investigate how repeated widefield optical stimulation affects neuronal firing over time, we analyzed the responses of MEA-recorded cultures to trains of light pulses delivered at 0.1, 0.2, 0.5, 1.0, and 2.0 Hz (see section \ref{sec:WF_exp}).\\
The spike trains from each MEA electrode were analyzed by separating the activity evoked by the stimulus from the ongoing "spontaneous" residual activity occurring between consecutive stimuli, to see if there could be a difference between them. To this end, the stimulation period $T$, defined as the temporal distance between the onset times of two consecutive light stimuli, was divided into two distinct time windows (blue box in figure \ref{fig:prima_figura}(C), table \ref{tab1}):
\begin{itemize}
    \item the \textit{stimulus response time window} $T_r$, with fixed width of $250$\,ms, starting at the onset of each stimulus. This window therefore includes both the spikes occurring during the $40$\,ms stimulation pulse itself and those within the following $210$\,ms after the stimulus offset;
    \item the \textit{baseline time window} $T_b$, extending from the end of $T_r$ for stimulus $n$ to the onset of the following stimulus $n+1$. The duration of $T_b$ is therefore variable, equal to $T - T_r$ and depending on the frequency $f_0$.
\end{itemize}
For each electrode $ch$ and stimulus $n$, two spike rates (section \ref{sec:rate}) were computed:
\begin{itemize}
    \item spike rate within $T_r$:
        \begin{equation}\label{eq:spike_rate_Tr_def} 
            \text{SR}_{\text{T}_\text{r}}(ch,n) = \frac{\# \,\text{spikes in }T_{\text{r}}}{T_{\text{r}}}
        \end{equation} 
    \item spike rate within $T_b$:
        \begin{equation}\label{eq:spike_rate_Tb_def} 
            \text{SR}_{\text{T}_\text{b}}(ch,n) = \frac{\# \,\text{spikes in }T_{\text{b}}}{T_{\text{b}}}
        \end{equation} 
\end{itemize}
Both rates are expressed in spikes per second. This analysis is done for all the $200$ stimuli (ending with $200$ values of spike rate for each channel of each analyzed MEA chip).
\\
To reduce trial-to-trial variability and highlight the underlying temporal evolution of the firing rate, the raw traces were smoothed using a moving-average filter with an adaptive window size. The window length was determined as a trade-off between: (i) local firing-rate variability, estimated from changes between consecutive pulses, (ii) the total stimulation duration relative to the stimulation period $T$. To prevent under-smoothing or excessive loss of temporal detail, the window size was constrained between 3 and 20 pulses.
\\
The results can be visualized for each electrode, during $T_r$ and $T_b$, for all the 5 frequencies $f_0$. Two examples can be seen in figure \ref{fig:results_teststimulus}(A) and (B). Each plot displays: the raw spike rate values (grey markers), the smoothed curve (blue line), and the pulse number on the x-axis. A clear frequency-dependent pattern emerges: at low stimulation frequencies ($0.1$ and $0.2$\,Hz), no electrode shows significant variations in firing rate during either $T_r$ or $T_b$, and the activity remains essentially constant over time (figure \ref{fig:results_teststimulus}(A)). At $0.5$\,Hz, in all the cultures, a subset of electrodes begins to exhibit a gradual, sigmoid-like decrease in firing rate. This behavior becomes more pronounced at $1.0$\,Hz and is consistently observed at $2.0$\,Hz across multiple electrodes (figure \ref{fig:results_teststimulus}(B)).
\\
This result is consistent with the known photocycle kinetics of Channelrhodopsin-2 (ChR2), which critically shape its ability to drive neuronal activity. Photon absorption triggers isomerization of the all-trans retinal chromophore, rapidly opening the channel with sub-millisecond kinetics \cite{nikolic2009photocycles}. Under whole-cell voltage clamp, a sustained blue-light stimulus evokes a characteristic photocurrent composed of an initial transient peak, a steady-state plateau, and a decay after light offset \cite{nikolic2009photocycles}. Under strong or repeated stimulation, photocurrents can decline by up to $\sim$80\% from peak to plateau, limiting sustained depolarization \cite{lin2011user}. The peak–plateau behavior reflects intrinsic ChR2 desensitization, whereby a substantial fraction of channels enter non-conducting or adapted states during continued illumination \cite{lin2011user}. Consistent with this, ChR2 photocurrents may require $5$-$25$\,s in darkness to fully recover following strong stimulation \cite{lin2011user}. Consequently, repeated light pulses delivered within the recovery time-window activate a reduced pool of available channels, producing progressively smaller photocurrents. These kinetics impose a strong frequency dependence on ChR2 efficacy: low-frequency stimulation permits near-complete recovery and reliable spike generation, whereas higher frequencies lead to reduced photocurrents and spike rates due to incomplete recovery \cite{lin2011user,stefanescu2013computational}. Computational studies confirm that such failures arise from photocycle kinetics rather than neuronal refractoriness, with four-state models accurately reproducing spike dropout at high stimulation rates \cite{stefanescu2013computational}. In our experimental findings, for low stimulation frequencies ($0.1$-$0.2$\,Hz) the long inter-pulse intervals ($5$-$10$\,s) are comparable to the recovery time constant of ChR2, allowing almost complete recovery of the channels from their desensitized states and resulting in stable, reproducible firing across successive pulses.
\\
To summarize the dynamics of neuronal adaptation during the stimulation train, quantitative parameters can be extracted by a sigmoidal fitting of the spike rate curves (red lines in figures \ref{fig:results_teststimulus}(A) and (B)). The model employed is the following:
\begin{equation}\label{eq:sigmoid_def} 
    y(x) = L + \frac{U - L}{1 + e^{k(x - t_0)}}
\end{equation}
where $L$ is the minimum observed firing rate, $U$ is the maximum, $t_0$ is the midpoint (pulse number at half-variation), and $k$ is the decay rate parameter, quantifying the rate of change. For each selected channel, the smoothed spike-rate profiles are fitted with this model using nonlinear least-squares regression (red lines in figure \ref{fig:results_teststimulus}(A)-(B)). The minimum and maximum are fixed to the values observed in the smoothed data, while $k$ and $t_0$ remain free parameters. This constraint reduces the degrees of freedom, improving fit stability and robustness, and ensures consistency with the visual trend of the data. A derived parameter, $\Delta = U-L$, quantifies the firing-rate decrease. In this way, the triplet $(k, t_0, \Delta)$ provides a compact yet informative description of the response dynamics.
\\
To isolate genuine spike rate's adaptation, only curves in which the maximum precedes the minimum are retained; fits violating this condition are discarded. Moreover, cases with $\Delta \le 10^{-3}\,s^{-1}$ are excluded from the analysis, as they likely reflect noise-driven fluctuations rather than meaningful changes in activity. The uncertainty of $\Delta$ is estimated by computing the local variability near the extrema (standard deviation in small windows around the maximum and minimum) and combining them. Confidence intervals for $k$ and $t_0$ are derived from the regression Jacobian.
Eventually, the fit parameters from different experiments are averaged. What emerges is that:
\begin{itemize}
    \item the slope parameter $k$ exhibits a strong dependence on stimulation frequency, rising steeply from nearly zero at low frequencies to about $0.05\,\text{s}^{-1}$ at $2.0$\,Hz (figure \ref{fig:results_teststimulus}(C)). This indicates that, as the stimulation rate increases, the transition from maximum to minimum firing rate becomes faster $-$ i.e., neuronal adaptation develops more abruptly;
    \item $\Delta$ values (representing only a negative change in spike rate) increase with frequency $f_0$. This trend indicates that higher stimulation frequencies induce larger variations in firing rate, reflecting a stronger modulation of neuronal excitability. In figure \ref{fig:results_teststimulus}(D), data corresponding to stimulation frequencies $0.1$ and $0.2$\,Hz are omitted, as the firing rate remains essentially constant over time at these frequencies, producing problems in the fitting process. In the absence of a measurable decay, the spike-rate profiles cannot be reliably described by a sigmoidal function, and the fitted parameters would not carry meaningful physiological information;
    \item $t_0$ results were excluded from our analysis since they did not show significant variations with stimulation frequency.
\end{itemize}
For both $T_r$ and $T_b$, the average values of the decay parameter $k$ at $0.1$ and $0.2$\,Hz are close to zero ($\sim 0.001\,\text{s}^{-1}$ and $\sim 0.002\,\text{s}^{-1}$, respectively), indicating the absence of appreciable firing-rate adaptation, in comparison with frequencies $\ge 0.5$\,Hz. 
Consistently, none of the analyzed electrodes exhibited detectable firing-rate variations at $0.2$\,Hz across all the cultures. Based on these observations, $0.2$\,Hz was selected as the test-stimulus frequency $f_{TS}$ for the LTP experiments. 
This choice represents an optimal compromise between response stability and experimental duration, allowing for shorter stimulation protocols compared to $0.1$\,Hz while preserving stable neuronal activity.

\begin{figure*}[!htbp]
\centering
\fbox{
  \begin{minipage}{1.0\linewidth}
    \centering
    \includegraphics[width=1.3\linewidth, angle=90]{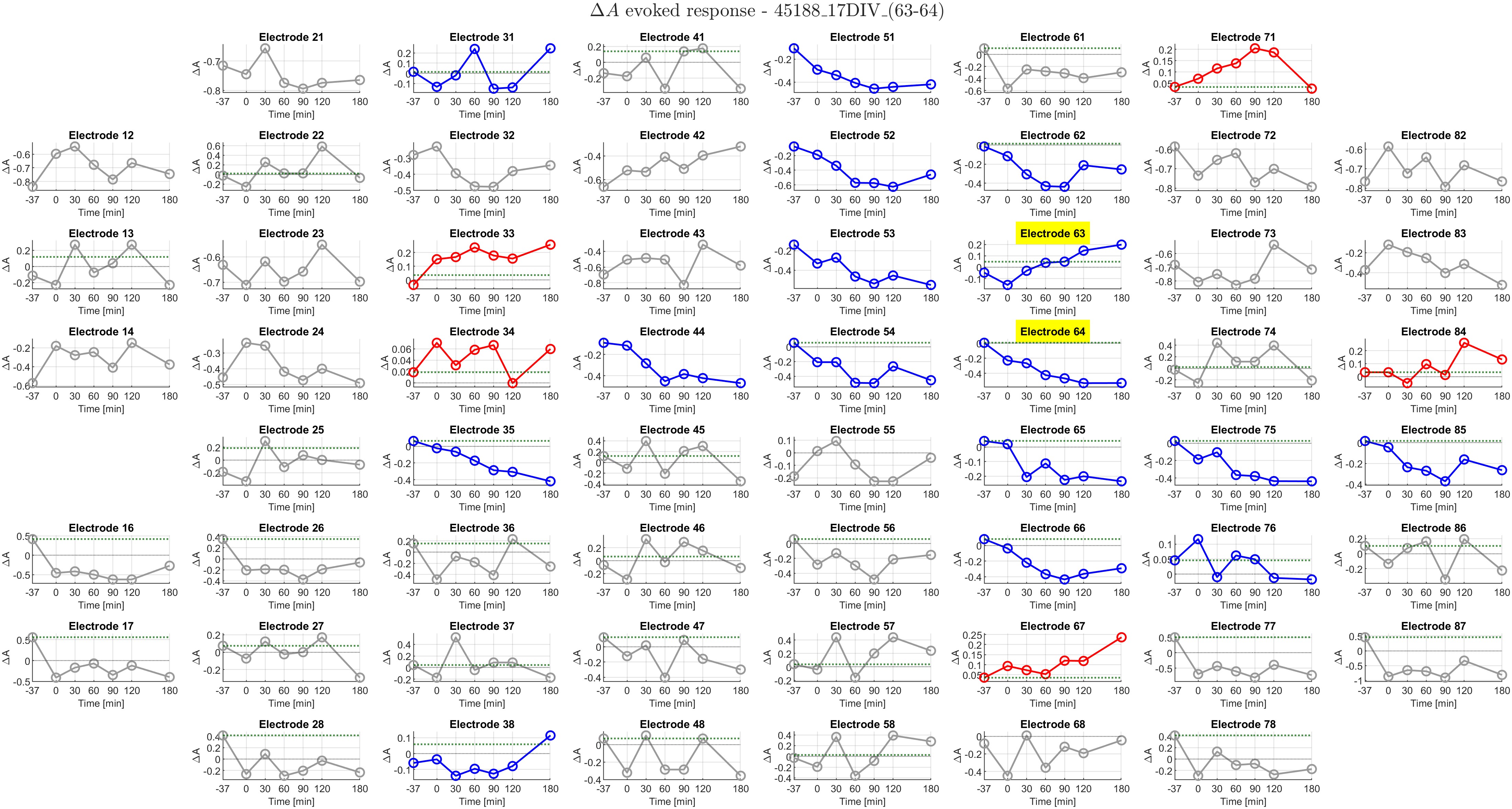}

    \caption{$\Delta A$ as a function of time for a single LTP-dataset. The plot shows the evolution of $\Delta A$ at different time steps for one of the 7 datasets. The x-axis shows all the 6 post-tetanus times plus the additional ‘baseline time' $-37$\,min (this is the moment at which \textit{Control 2} stimulation started).
    Values are plotted in grey, red, or blue according to the classification explained in the text. The two electrodes chosen for tetanic stimulation are highlighted in yellow. The $S$ value for each electrode is indicated by a green dotted line (in subplots where the $S$ level falls outside the y-axis range defined by the data points, the line is not displayed).}
    \label{fig:MEA_map_LTP}
  \end{minipage}
}
\end{figure*}

\subsection{Results for LTP}\label{sec:LTP_results}
After determining the test-stimulus frequency $f_{TS}$, we proceeded to investigate long-term potentiation on neuronal cultures.
To this end, the LTP stimulation protocol described in section \ref{sec:LTP_exp} was applied using $f_0=f_{TS}=0.2$\,Hz as the test-stimulus frequency before and after tetanic stimulation.
Upon completion of the experiments, spike data were acquired from all electrodes and subsequently analyzed to evaluate both the magnitude and the temporal evolution of post-tetanic changes in neuronal responses.
\\
To visualize how post-tetanic responses evolve over time and across the MEA, we generated spatial maps arranged according to the $8\times8$ electrode layout (see an example in figure \ref{fig:MEA_map_LTP}). In the maps, each subplot corresponds to a single electrode and displays the temporal evolution of $\Delta A$ (see section \ref{sec:efficacy}) across the six post-tetanus time points. In this representation, both the spatial heterogeneity of potentiation and its temporal persistence can be readily appreciated. An additional point (at time $-37$\,min) $-$ evaluated as $[Area_{C2}(ch) - Area_{BC}(ch)]/Area_{BC}(ch)$ $-$ is added in each subplot to visualize the variation of $\Delta A$ values with respect to baseline activity.
\\
A critical aspect of this analysis is the ability to distinguish genuine LTP-related changes from fluctuations arising from baseline variability. To this end, we introduced a channel-specific \textit{stability} parameter $S$, which quantifies the intrinsic variability between the two control recordings for each electrode.
The stability parameter is defined as:
\begin{equation}\label{eq:S_stability} 
    S(ch) = \left|\frac{Area_{C2}(ch) - Area_{C1}(ch)}{Area_{C2}(ch) + Area_{C1}(ch)}\right|
\end{equation} 
The parameter $S$ therefore provides a reference scale for assessing whether post-tetanic changes exceed the level of variability expected under baseline conditions. In practice, post-induction responses must exceed a multiple of $S$ to be considered meaningful (see section \ref{sec:PI}). 
\\
Based on the combined information provided by baseline activity, stability, and post-tetanic response amplitude, each electrode was classified into one of three categories. These categories are displayed using different colors in the $\Delta A$ plots (figure \ref{fig:MEA_map_LTP}):
\begin{enumerate}
    \item \textit{Red electrodes} (significant potentiation): An electrode is highlighted in red when three conditions are simultaneously satisfied:
        \begin{enumerate}[label=\alph*)]
            \item the baseline response is sufficiently strong ($Area_{BC}\ge 0.5$), ensuring that the channel carries reliable activity;
            \item the variability between the two control recordings is low ($S \le 0.2$), which means that the electrode’s baseline is stable;
            \item the average of the $6$ post-tetanic responses shows a value of efficacy $\Delta A \ge 1.2\times S$, indicating that the change of PSTH area is greater than the one expected from natural variability.
        \end{enumerate}  
        Red traces, therefore, represent electrodes with robust baseline activity, stable control conditions, and a consistent potentiation effect.
        
    \item \textit{Blue electrodes} (stable but not potentiated):
    these are electrodes that satisfy the conditions a) and b) $-$ strong baseline and low variability $-$ but do not show a sufficient increase beyond the variability threshold. We have considered them as reliable, but they are not exhibiting significant potentiation under criterion c).
    
    \item \textit{Grey electrodes} (non-reliable):
    electrodes that fail the baseline activity threshold or show excessive variability between the two controls. These channels were excluded from our analysis since changes in their response cannot be distinguished from noise or instability.
\end{enumerate}

\begin{figure*}[!htbp]
    \centering
    \fbox{
      \begin{minipage}{1.0\linewidth}
        \centering
        \includegraphics[width=\linewidth]{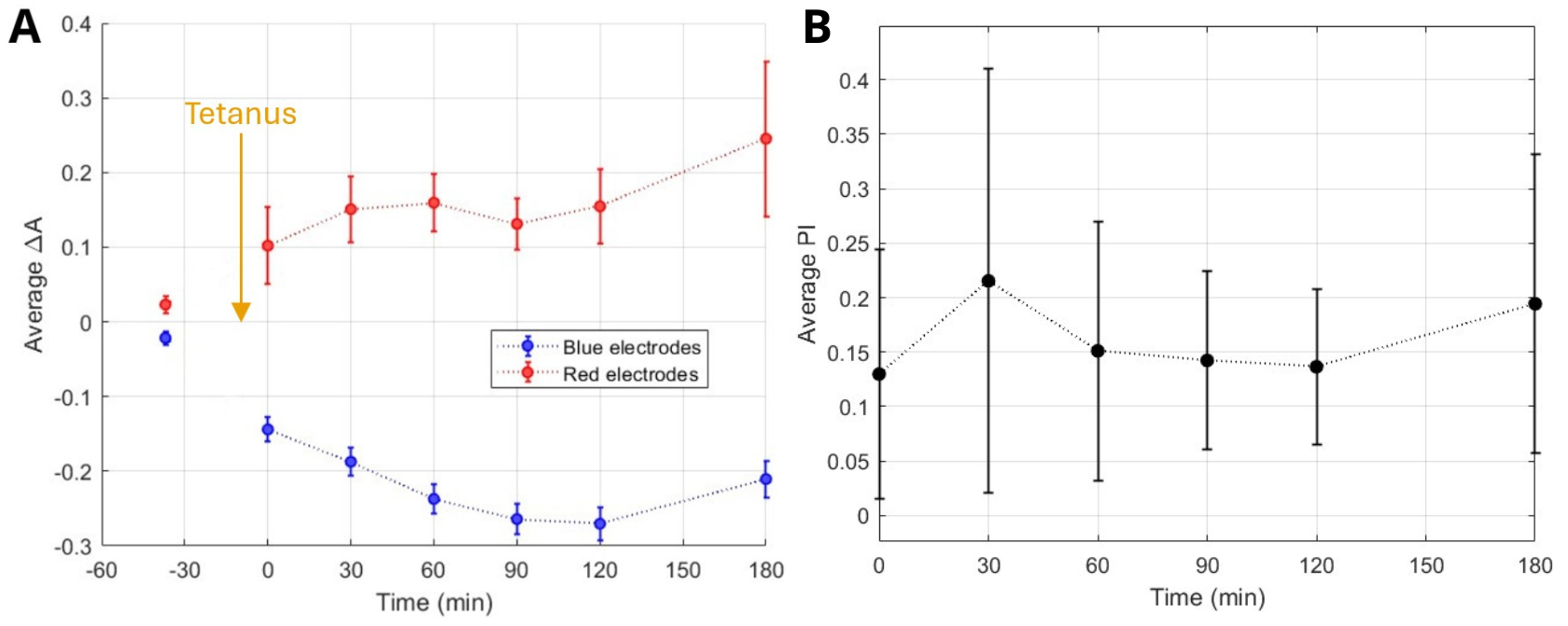}
    
        \caption{(A): The plot shows the mean value of $\Delta A$ computed across all electrodes classified as red and blue, at the different times after the LTP induction protocol. The additional data at -$37$ min correspond to $\Delta A$ evaluated for \textit{Control 2}, to give a baseline reference. The yellow arrow indicates the moment of the start of the tetanic stimulation. (B): Average Potentiation Index (PI) as a function of time post-LTP, computed across the 7 independent datasets. Each point represents the mean PI value at a given time, with error bars indicating the standard deviation across experiments.}
        \label{fig:risultati_LTP}
      \end{minipage}
    }   
\end{figure*}

In addition to the MEA maps, a plot presenting the average behavior of all red and blue electrodes across all experiments was also produced (figure \ref{fig:risultati_LTP}(A)).
A clear divergence between the groups of electrodes emerges following tetanic stimulation. Electrodes classified as potentiated (red) exhibit an increase in $\Delta A$ over time, indicating a gradual and sustained enhancement of the evoked response. This behavior is consistent with the induction of long-term potentiation in these channels. In contrast, stable but non-potentiated electrodes (blue) show a decrease in $\Delta A$, reaching negative values within the first hour post-stimulation. This trend may reflect compensatory mechanisms, such as homeostatic regulation or redistribution of network activity following localized potentiation \cite{turrigiano2004homeostatic,turrigiano2012homeostatic}. 
Together, these results suggest that the stimulation paradigm induces lasting potentiation in some electrodes, whilst the opposite modulation between the two electrode groups supports the presence of coordinated, network-level plasticity dynamics following LTP induction. This could be a starting point for further investigation on the distribution of potentiation spreading across neuronal networks\cite{auslender2025decoding}. No variations of the spike rate were observed for all the spontaneous activity recording reported in section \ref{sec:LTP_exp}. 
\\
To quantitatively assess the overall strength and persistence of light-induced potentiation across the MEA channels, the PI index (see section \ref{sec:PI}) is computed at every post-tetanus time $t$. This index thus provides a normalized, network-level measure of the fraction of electrodes exhibiting potentiation over time, independent of culture size or electrode count \cite{chiappalone2008network}. 
After computing the PI for each dataset, the mean and standard deviation were evaluated across experiments for each time point. The resulting summary plot is shown in figure \ref{fig:risultati_LTP}(B): each point in the graph corresponds to the mean PI at a given post-tetanus time, with error bars indicating the standard deviation. This global representation enables an intuitive assessment of how the average degree of potentiation evolves in time across all analyzed recordings, providing an integrated measure of network-level strengthening following LTP induction. 
\\
Immediately after tetanus' conclusion ($t=0$), the average PI shows a modest increase, indicating that a subset of electrodes responds with enhanced evoked activity. The PI reaches its maximum approximately $30$\,min post-stimulation, corresponding to the peak of network-level potentiation. After this initial rise, the PI gradually declines and stabilizes around $0.15$-$0.20$, suggesting that although some degree of potentiation decays over time, a fraction of electrodes ($\sim 20 \%$) remains potentiated for at least three hours after the tetanic stimulus. The relatively large variability observed at early time points likely reflects differences across cultures in the spatial extent and strength of induced potentiation.
\\
Overall, these results confirm that the optogenetic stimulation paradigm reliably induces long-lasting potentiation across multiple experiments, with a stable portion of the network maintaining enhanced responsiveness over extended timescales.

\section*{Conclusions}\label{sec:conclusions}
In this work, we presented a study of optogenetically induced long-term potentiation on neuronal cultures, with activity recorded by means of microelectrode arrays. We used a widefield test stimulus to probe the responses of the largest possible number of electrodes within the illuminated area. By explicitly characterizing the frequency-dependent adaptation of optogenetically evoked responses, we identified a low-frequency ($0.2$\,Hz) widefield test-stimulus that reliably probes network activity while minimizing ChR2 desensitization. Building on this optimized protocol, spatially confined tetanic optical stimulation induced robust and localized potentiation, persisting for at least $3$ hours after induction. The combination of DLP-based patterned illumination, MEA recordings, and quantitative PSTH-based metrics enabled assessment of network potentiation. Overall, this study shows that optogenetic stimulation, when combined with carefully designed control paradigms and analysis methods, provides a powerful and reproducible approach for probing activity-dependent plasticity at the network scale, paving the way for controlled investigations of learning-related mechanisms in simplified neuronal systems.

\suppdata{The data that support the findings of this study are available on: \url{https://doi.org/10.5281/zenodo.14363732}.
The MATLAB codes developed for the analysis are available at: \url{https://github.com/MatteoDominici/matlab-codes-IOP-2026}.}

\ack{This work was financed by the European Union - NextGenerationEU - National Recovery and Resilience Plan (NRRP) - Mission 4 Component 2 Investment 1.2 – “Funding projects presented by young researchers” MSCA PNRR Young Researchers, “CIRCUS project” - MSCA20240000106 - CUP E63C25000820007.\\We thank Master student Aleyna Habesoglu for valuable
help in the preparation of MEA chips and samples.}

\bibliographystyle{vancouver}

\end{document}